\documentclass[prd,twocolumn,showpacs,preprintnumbers,amsmath,amssymb]{revtex4}

\usepackage{amsmath}
\usepackage{latexsym}
\def\[{\left\lbrack}
\def\]{\right\rbrack}

\def\({\left(}
\def\){\right)}
\newcommand{\be}{\begin{equation}}
\newcommand{\ee}{\end{equation}}
\newcommand{\ea}{\end{eqnarray}}
\newcommand{\ba}{\begin{eqnarray}}

\begin{document}

\title{Duality considerations about the Maxwell-Podolsky theory through \\ the symplectic embedding formalism and spectrum analysis}

%\author{E. M. C. Abreu$^a$\footnote{\sf E-mail: evertonabreu@ufrrj.br}, A. C. R. Mendes$^b$\footnote{\sf E-mail: albert@ufv.br},C. Neves$^c$\footnote{\sf E-mail: clifford.neves@gmail.com}, W. Oliveira$^d$\footnote{\sf E-mail: wilson@fisica.ufjf.br}, C. Wotzasek$^d$\footnote{\sf E-mail: wotsazek@if.ufrj.br} and L. M. V. Xavier}
\author{E. M. C. Abreu$^a$\footnote{\sf E-mail: evertonabreu@ufrrj.br}, A. C. R. Mendes$^b$\footnote{\sf E-mail: albert@ufv.br}, C. Neves$^c$\footnote{\sf E-mail: clifford.neves@gmail.com}, W. Oliveira$^d$\footnote{\sf E-mail: wilson@fisica.ufjf.br}, C. Wotzasek$^d$\footnote{\sf E-mail: wotsazek@if.ufrj.br} and L. M. V. Xavier$^d$}  
\affiliation{${}^{a}$Grupo de F\' isica Te\'orica e Matem\'atica F\' isica,
Departamento de F\'{\i}sica, Universidade Federal Rural do Rio de Janeiro\\
BR 465-07, 23890-000, Serop\'edica, Rio de Janeiro, Brazil\\
${}^{b}$Campus Rio Parana\' iba, Universidade Federal de Vi\c{c}osa,\\
BR 354 - km 310, 38810-000, Rio Parana\' iba, Minas Gerais, Brazil\\
${}^{c}$Departamento de Matem\'atica e Computa\c{c}\~ao, Universidade do Estado do Rio de Janeiro\\
Rodovia Presidente Dutra, km 298, 27537-000, Resende, Rio de Janeiro, Brazil\\
${}^{d}$Departamento de F\'{\i}sica,ICE, Universidade Federal de Juiz de Fora,\\
36036-330, Juiz de Fora, MG, Brazil\\
${}^{e}$Instituto de F\'\i sica, Universidade Federal do Rio de
Janeiro, 21945, Rio de Janeiro, Brazil\\
\bigskip
\today\\}

\begin{abstract}
We find the dual equivalent (gauge invariant) version of the Maxwell theory in $D=4$ with a Proca-like mass term by using the symplectic embedding method. The dual theory obtained (Maxwell-Podolsky) includes a higher-order derivative term and preserve the gauge symmetry.  We also furnish an investigation of the pole structure of the vector propagator by the residue matrix  which considers the eventual existence of the negative-norm of the theory.
\end{abstract}

\maketitle

\section{Introduction}

The interest in the dual mapping between two different theories that show the same physical properties has been increased in the last years motivated by its success in both field theory as well as string theory \cite{az}.  At the same time, the interest in he study of theories involving higher-order derivatives is by now well appreciated and remains intense.  Within the context of Maxwell theory, generalizations involving higher-order derivatives can be found in \cite{BI,Proca,Podol}.

We know that in $D=3+1$ massive gauge fields are introduced by the Higgs Mechanism through the spontaneous breaking of symmetry. However, in the Abelian case, it is possible to introduce a  massive gauge field without  having to accomplish a spontaneous breaking of symmetry. So,  the gauge symmetry does not forbid the appearance of the massive gauge fields. We can fall back to a topological term of mass like Chern-Simons \cite{CS}, that in  $D=2+1$ preserve the gauge and Lorentz symmetries. However, in $D=3+1$ the Lorentz symmetry is broken for the Chern-Simons theory.

There are different ways of extending electrodynamics, trying to smooth infrared or ultraviolet singularities that appear at large or short distances. One is the Born-Infeld \cite{BI} type of generalization, that involves nonlinear extension. The other includes the generalizations introduced by Proca \cite{Proca}, that involves the addition of a mass term for the vector field, which was introduced to smooth infrared singularities. But, the gauge symmetry is clearly lost, due to the term $A_{\mu}A^{\mu}$. Another way is through the work of Podolsky \cite{Podol}, that involves higher-order derivatives and was introduced to smooth ultraviolet singularities. The Podolsky theory propagates a massive photon and the Lorentz and gauge symmetries are not spontaneously broken. Besides, it has a fundamental role in the discussion about the compatibility of the magnetic monopole and massive photons. 

Working in $D=3$, the authors in \cite{Bazeia} presented the dual mapping of the Maxwell-Chern-Simons-Proca (MCS-Proca) with a MCS-Podolsky term  through the iterative gauge embedding procedure \cite{Anac}. It was introduced a convenient parameter $a$ in the Maxwell term.  When $a=0$ we have a MCS-Proca, that is the standard self-dual model \cite{Town}, and when $a=1$ it was obtained an extended CS-Podolsky model.   The objective of the introduction of this parameter is to construct a master action, from which one can obtain both the model and its dual partner.  But one can question if this dual partner is unique.  We believe that the symplectic embedding technique can furnish a whole family of dual models because of the arbitrariness of the zero-mode which is also the generator of the infinitesimal gauge transformations, as we will see below.

Consequently, the purpose of this paper is to work out the dual aspects of electrodynamics that appears in the Proca and Podolsky models by the symplectic embedding  method \cite{NPB}, that has been very efficient to reveal a desired hidden symmetry and a dual partner of the model \cite{NPB,JPA,amnot,MPLA}.  We analyze the residue matrix at each pole of the propagators. The poles furnishes a relation between energy and momentum which can be associated to a massive particle. The residue matrix gives information about the degree of freedom of the partner polarization, where physical states are partnered with positive eigenvalues of the residue matrix at each pole \cite{helayel}. The existence of negative eigenvalues shows that in these situations nonphysical states that correspond to negative norm particle states (ghosts) are introduced.

The symplectic embedding method (SEM) \cite{NPB} is not affected by ambiguity problems. It has the great advantage of being a simple and direct way of choosing the infinitesimal gauge generators of the built gauge theory. As a consequence, we have the freedom to choose the content of the embedded symmetry according to our necessities. This feature makes possible a greater control over the final Lagrangian. This method can avoid the introduction of infinite terms in the Hamiltonian of embedded non-commutative and non-Abelian theories. This can be accomplished because the infinitesimal gauge generators does not result from previous unclear choices. Another advantage related  is the possibility of doing a kind of general embedding.  In other words we can say that, instead of choosing the gauge generators at the beginning, one can leave some unfixed parameters with the aim of fixing them later, when the final Lagrangian has being achieved. 
Although a faster way to obtain the final theory is through the fixing of such parameters, it is more interesting in order to obtain knowledge of the in question.  Also, it is helpful if we want to know its hidden symmetries, but some aspects of the Lagrangian are required.
The embedding approach is not dependent on any undetermined constraint structure and it also works for unconstrained systems. This is different from all the existent
embedding techniques that we can use to convert \cite{Bata,Wotz}, to project \cite{Vyth} or to reorder \cite{Mitra} the existence of the second-class constraints into a first-class system. This technique on the other hand only deals with the symplectic structure of the theory,  so this embedding structure does not rely on any pre-existent constrained structure.

In order to make this paper self-sustained, in the next section we present a brief review of the dual embedding formalism. In the section \ref{s3}, the SEM will be used in the Proca model to construct a massive gauge invariant theory. In the section \ref{s4}  we present an investigation of the structure of the gauge propagator. Finally, the conclusions are accomplished in the last section, as usual. Our metric tensor has signature $(+---)$ and we use natural units.

\section{The  dual embedding formalism}\label{s2}

This technique follows the Faddeev-Shatashivilli's
suggestion \cite{FS} and is set up on a contemporary framework to handle constrained
models, i.e., the symplectic formalism \cite{FJ}.  
In the following lines, we try to keep this paper self-sustained reviewing the main steps of the dual embedding formalism.  We will follow closely the ideas contained in 
\cite{amnot}.

Let us consider a general noninvariant mechanical model whose dynamics is governed by a Lagrangian
${\cal L}(a_i,\dot a_i,t)$, (with $i=1,2,\dots,N$), where $a_i$ and $\dot a_i$
are the space and velocities
variables, respectively. Notice that this model does not result in the loss of
generality nor physical content. Following the symplectic method the zeroth-iterative
first-order Lagrangian one-form is written as
 \begin{equation}
\label{2000}
{\cal L}^{(0)}dt = A^{(0)}_\theta d\xi^{(0)\theta} - V^{(0)}(\xi)dt,
\end{equation}
and the symplectic variables are
\be
\xi^{(0)\theta} =  \left\{ \begin{array}{ll}
                               a_i, & \mbox{with $\theta=1,2,\dots,N $} \\
                               p_i, & \mbox{with $\theta=N + 1,N + 2,\dots,2N ,$}
                           \end{array}
                     \right.
\ee
where $A^{(0)}_\theta$ are the canonical momenta and $V^{(0)}$ is the symplectic potential. From the Euler-Lagrange equations of motion, the symplectic tensor is obtained as
\begin{eqnarray}
\label{2010}
f^{(0)}_{\theta\beta} = {\partial A^{(0)}_\beta\over \partial \xi^{(0)\theta}}
-{\partial A^{(0)}_\theta\over \partial \xi^{(0)\beta}}.
\end{eqnarray}
If the two-form $f \equiv \frac{1}{2}f_{\theta\beta}d\xi^\theta \wedge d\xi^\beta$ is singular, the symplectic matrix (\ref{2010}) has a zero-mode $(\nu^{(0)})$ that generates a new constraint when contracted with the gradient of the symplectic potential,
\begin{equation}
\label{2020}
\Omega^{(0)} = \nu^{(0)\theta}\frac{\partial V^{(0)}}{\partial\xi^{(0)\theta}}.
\end{equation}
This constraint is introduced into the zeroth-iterative Lagrangian one-form equation (\ref{2000}) through a Lagrange multiplier $\eta$, generating the next one
\begin{eqnarray}
\label{2030}
{\cal L}^{(1)}dt &=& A^{(0)}_\theta d\xi^{(0)\theta} + d\eta\Omega^{(0)}- V^{(0)}(\xi)dt,\nonumber\\
&=& A^{(1)}_\gamma d\xi^{(1)\gamma} - V^{(1)}(\xi)dt,\end{eqnarray}
with $\gamma=1,2,\dots,(2N + 1)$ and
\begin{eqnarray}
\label{2040}
V^{(1)}&=&V^{(0)}|_{\Omega^{(0)}= 0},\nonumber\\
\xi^{(1)_\gamma} &=& (\xi^{(0)\theta},\eta),\\
A^{(1)}_\gamma &=&(A^{(0)}_\theta, \Omega^{(0)}).\nonumber
\end{eqnarray}
As a consequence, the first-iterative symplectic tensor is computed as
\begin{eqnarray}
\label{2050}
f^{(1)}_{\gamma\beta} = {\partial A^{(1)}_\beta\over \partial \xi^{(1)\gamma}}
-{\partial A^{(1)}_\gamma\over \partial \xi^{(1)\beta}}.
\end{eqnarray}
If this tensor is nonsingular, the iterative process stops and the Dirac's brackets
 among the phase space variables are obtained from the inverse matrix
 $(f^{(1)}_{\gamma\beta})^{-1}$ and, consequently, the Hamilton equation of
 motion can be computed and solved, as discussed in \cite{gotay}. It is well known
 that a physical system can be described at least classically in terms of a symplectic
 manifold $M$. From a physical point of view, $M$ is the phase space of the system while
 a nondegenerate closed 2-form $f$ can be identified as being the Poisson bracket. The
 dynamics of the system is  determined just specifying a real-valued function (Hamiltonian)
$H$ on the phase space, {\it i.e.}, one of these real-valued function
solves the Hamilton equation, namely,
\be \label{2050a1}
\iota(X)f=dH, \ee
and the classical dynamical trajectories of the
system in the phase space are obtained. It is important to mention
that if $f$ is nondegenerate, equation (\ref{2050a1}) has an unique
solution. The nondegeneracy of $f$ means that the linear map
$\flat:TM\rightarrow T^*M$ defined by $\flat(X):=\flat(X)f$ is an
isomorphism, due to this, the equation (\ref{2050a1}) is solved uniquely
for any Hamiltonian $(X=\flat^{-1}(dH))$. On the contrary, the
tensor has a zero-mode and a new constraint arises, indicating
that the iterative process goes on until the symplectic matrix
becomes nonsingular or singular. If this matrix is nonsingular,
the Dirac's brackets will be determined. In Ref. \cite{gotay}, the
authors consider in detail the case when $f$ is degenerate. The main idea of this embedding formalism is to introduce extra fields into the model in order to obstruct the solutions of the Hamiltonian equations of motion.
We introduce two arbitrary functions which are dependent on the original phase space and of WZ's variables, namely, $\Psi(a_i,p_i)$ and $G(a_i,p_i,\eta)$, into the first-order Lagrangian one-form as follows
\be
\label{2060a}
{\tilde{\cal L}}^{(0)}dt = A^{(0)}_\theta d\xi^{(0)\theta} + \Psi d\eta - {\tilde V}^{(0)}(\xi)dt,
\ee
with
\be
\label{2060b}
{\tilde V}^{(0)} = V^{(0)} + G(a_i,p_i,\eta),
\ee
where the arbitrary function $G(a_i,p_i,\eta)$ is expressed as an expansion in terms of the WZ field, given by
\begin{equation}
\label{2060}
G(a_i,p_i,\eta)=\sum_{n=1}^\infty{\cal G}^{(n)}(a_i,p_i,\eta),\,\,\,\,\,\,\,{\cal G}^{(n)}(a_i,p_i,\eta)\sim\eta^n\,,
\end{equation}
and satisfies the following boundary condition
\begin{eqnarray}
\label{2070}
G(a_i,p_i,\eta=0) = 0.
\end{eqnarray}
The symplectic variables were extended to also contain the WZ variable $\tilde\xi^{(0)\tilde\theta} = (\xi^{(0)\theta},\eta)$ (with ${\tilde\theta}=1,2,\dots,2N+1$) and the first-iterative symplectic potential becomes
\begin{equation}
\label{2075}
{\tilde V}^{(0)}(a_i,p_i,\eta) = V^{(0)}(a_i,p_i) + \sum_{n=1}^\infty{\cal G}^{(n)}(a_i,p_i,\eta).
\end{equation}
In this context, the new canonical momenta are
\be
{\tilde A}_{\tilde\theta}^{(0)} = \left\{\begin{array}{ll}
                                  A_{\theta}^{(0)}, & \mbox{with $\tilde\theta$ =1,2,\dots,2N}\\
                                  \Psi, & \mbox{with ${\tilde\theta}$= 2N + 1}
                                    \end{array}
                                  \right.
\ee
and the new symplectic tensor, given by
\begin{equation}
{\tilde f}_{\tilde\theta\tilde\beta}^{(0)} = \frac {\partial {\tilde A}_{\tilde\beta}^{(0)}}{\partial \tilde\xi^{(0)\tilde\theta}} - \frac {\partial {\tilde A}_{\tilde\theta}^{(0)}}{\partial \tilde\xi^{(0)\tilde\beta}},
\end{equation}
that is
\be
\label{2076b}
{\tilde f}_{\tilde\theta\tilde\beta}^{(0)} = 
\begin{pmatrix}
 { f}_{\theta\beta}^{(0)} & { f}_{\theta\eta}^{(0)}
\cr { f}_{\eta\beta}^{(0)} & 0
\end{pmatrix}.
\ee

To sum up we have two steps: the first one is addressed to compute $\Psi(a_i,p_i)$ while the second one is dedicated to the calculation of $G(a_i,p_i,\eta)$. In order to begin with the first step, we impose that this new symplectic tensor (${\tilde f}^{(0)}$) has a zero-mode $\tilde\nu$, consequently, we get the following condition
\begin{equation}
\label{2076}
\tilde\nu^{(0)\tilde\theta}{\tilde f}^{(0)}_{\tilde\theta\tilde\beta} = 0.
\end{equation}
At this point, $f$ becomes degenerate and, in consequence, we introduce an obstruction to solve, in an unique way, the Hamilton equation of motion given in equation (\ref{2050a1}). Assuming that the zero-mode $\tilde\nu^{(0)\tilde\theta}$ is
\begin{equation}
\label{2076a}
\tilde\nu^{(0)}=
\begin{pmatrix}
\mu^\theta & 1
\end{pmatrix},
\end{equation}
and using the relation given in (\ref{2076}) together with (\ref{2076b}), we get a set of equations, namely,
\be
\label{2076c}
\mu^\theta{ f}_{\theta\beta}^{(0)} + { f}_{\eta\beta}^{(0)} = 0,
\ee
where
\be
{ f}_{\eta\beta}^{(0)} =  \frac {\partial A_\beta^{(0)}}{\partial \eta} - \frac {\partial \Psi}{\partial \xi^{(0)\beta}}.
\ee
The matrix elements $\mu^\theta$ are chosen in order to disclose a desired gauge symmetry. Note that in this formalism the zero-mode $\tilde\nu^{(0)\tilde\theta}$ is the gauge symmetry generator. At this point, it is worth to mention that this characteristic is important because it opens up the possibility to disclose the desired hidden gauge symmetry from the noninvariant model. It awards to the symplectic embedding formalism some power to deal with noninvariant systems. From relation (\ref{2076}) some differential equations involving $\Psi(a_i,p_i)$ are obtained, equation (\ref{2076c}), and after a straightforward computation, $\Psi(a_i,p_i)$ can be determined.

In order to compute $G(a_i,p_i,\eta)$ in the second step, we impose that no more constraints arise from the contraction of the zero-mode $(\tilde\nu^{(0)\tilde\theta})$ with the gradient of the potential ${\tilde V}^{(0)}(a_i,p_i,\eta)$. This condition generates a general differential equation, which reads as
\begin{widetext}
\begin{eqnarray}
\label{2080}
\tilde\nu^{(0)\tilde\theta}\frac{\partial {\tilde V}^{(0)}(a_i,p_i,\eta)}{\partial{\tilde\xi}^{(0)\tilde\theta}}\,&=&\, 0,\\
\mu^\theta \frac{\partial {V}^{(0)}(a_i,p_i)}{\partial{\xi}^{(0)\theta}} + \mu^\theta \frac{\partial {\cal G}^{(1)}(a_i,p_i,\eta)}{\partial{\xi}^{(0)\theta}} 
\,+\, \mu^\theta\frac{\partial {\cal G}^{(2)}(a_i,p_i,\eta)}{\partial{\xi}^{(0)\theta}} &+& \dots \nonumber\\
\,+\,\frac{\partial {\cal G}^{(1)}(a_i,p_i,\eta)}{\partial\eta} + \frac{\partial {\cal G}^{(2)}(a_i,p_i,\eta)}{\partial\eta} &+& \dots = 0\;\;, 
\end{eqnarray}
\end{widetext}
that allows us to compute all correction terms ${\cal G}^{(n)}(a_i,p_i,\eta)$ in order of $\eta$. Note that this polynomial expansion in terms of $\eta$ is equal to zero, consequently, whole coefficients for each order in $\eta$ must be null identically. In view of this, each correction term in order of $\eta$ is determined. For a linear correction term, we have
\begin{equation}
\label{2090}
\mu^\theta\frac{\partial V^{(0)}(a_i,p_i)}{\partial\xi^{(0)\theta}} + \frac{\partial{\cal
 G}^{(1)}(a_i,p_i,\eta)}{\partial\eta} = 0.
\end{equation}
For a quadratic correction term, we get
\begin{equation}
\label{2095}
{\mu}^{\theta}\frac{\partial{\cal G}^{(1)}(a_i,p_i,\eta)}{\partial{\xi}^{(0)\theta}} + \frac{\partial{\cal G}^{(2)}(a_i,p_i,\eta)}{\partial\eta} = 0.
\end{equation}
From these equations, a recursive equation for $n\geq 2$ is proposed as
\begin{equation}
\label{2100}
{\mu}^{\theta}\frac{\partial {\cal G}^{(n - 1)}(a_i,p_i,\eta)}{\partial{\xi}^{(0)\theta}} + \frac{\partial{\cal
 G}^{(n)}(a_i,p_i,\eta)}{\partial\eta} = 0,
\end{equation}
that allows us to compute the remaining correction terms in order of $\eta$. This iterative process is successively repeated until (\ref{2080}) becomes identically null, consequently, the extra term $G(a_i,p_i,\eta)$ is obtained explicitly. Then, the gauge invariant Hamiltonian, identified as being the symplectic potential, is obtained as
\begin{equation}
\label{2110}
{\tilde{\cal  H}}(a_i,p_i,\eta) = V^{(0)}(a_i,p_i) + G(a_i,p_i,\eta),
\end{equation}
and the zero-mode ${\tilde\nu}^{(0)\tilde\theta}$ is identified as being the generator of an infinitesimal gauge transformation, given by
\begin{equation}
\label{2120}
\delta{\tilde\xi}^{\tilde\theta} = \varepsilon{\tilde\nu}^{(0)\tilde\theta},
\end{equation}
where $\varepsilon$ is an infinitesimal parameter.

\section{Maxwell-Podolsky from Maxwell-Proca}\label{s3}

%In this section, the Maxwell theory with a Proca-like mass term in $D=3+1$ will be analyzed from SEM.

In this section we follow the steps described in the last one in order to find a dual equivalent action to the original theory.  As we will see this dualization allows us to construct a mapping between a non gauge invariant model (Proca) and a gauge invariant one (Podolsky).

Let us consider the massive Lagrangian density in four dimensions  

\be
\label{01}
{\cal L}_{{\rm MProca}}=-\frac{1}{4}F_{\mu\nu}F^{\mu\nu}+\frac{\xi}{2}A_{\mu}A^{\mu},
\ee
where the canonical momenta are
\be
\label{02}
\pi_i = -F_{0i}.
\ee
 
Hence, we have the following Lagrangian,
\be
\label{03}
{\cal L}^{(0)}=\pi^{i}\dot{A}_{i}-V^{(0)}
\ee
where
\be\label{04}
V^{(0)}=\frac{1}{4}F_{ij}F^{ij}-\frac{\xi}{2}A_{\mu}A^{\mu}-\frac{1}{2}\pi^{i}\pi_{i}+\pi_{i}\partial^{i}A_{0}.
\ee

Considering a convenient zero-mode as,
\be\label{14}
\tilde\nu^{(0)}=\left(
\begin{array}{cccc}
\partial_{x}^{i} & 0 & \partial^{0}& -1
\end{array}\right)
\ee
and extending the phase space with the introduction of the WZ fields and the two arbitrary function $\psi\equiv\psi\left(A^{i},\psi^{i},A^{0},\eta \right)$ and $G\equiv G\left( A^{i},\psi^{i},A^{0},\eta \right)$ we change the Lagrangian (\ref{03}), namely,
\be\label{15}
\tilde{{\cal L}}^{(0)}=\pi^{i}\dot{A}_{i} +\psi \dot \eta- \tilde{V}^{(0)}
\ee
where the symplectic potential is
\be\label{16}
\tilde{V}^{(0)}=\frac{1}{4}F_{ij}F^{ij}-\frac{\xi}{2}A_{\mu}A^{\mu}-\frac{1}{2}\pi^{i}\pi_{i}+\pi_{i}\partial^{i}A_{0} + G.
\ee
The function $G$ is expressed as
\be\label{17}
G\left( A^{i},\psi^{i},A^{0},\eta \right)=\sum_{n=0}^{\infty}{\cal G}^{n} \;\;{\rm with}\;\; {\cal G}^{n} \propto \eta^{n}.
\ee

Now, contracting the zero-mode (\ref{14}) with the symplectic matrix of the Lagrangian (\ref{15}), and after a straightforward computation, we get
\be\label{18}
\psi=-\partial_i \pi^{i}.
\ee
Then, the Lagrangian becomes
\be\label{19}
\tilde{{\cal L}}^{(0)}=\pi^{i}\dot{A}_{i} -\partial_i \pi^{i}\dot \eta- \tilde{V}^{(0)}.
\ee

Contracting the zero-mode (\ref{14}) with the symplectic potential we have that,
\ba\label{20}
& &\int dy \left(\left[ -\partial_{y}^{i} F_{ij}({\bf y}) -{\xi} A_{j}({\bf y})\right]\partial_{x}^{j}\delta({\bf x} -{\bf y}) \right. \nonumber \\
&+& \left. \left[-\partial_{y}^{i}\pi_{i}({\bf y}) -{\xi} A_{0}({\bf y}) \right]\partial_{x}^{0}\delta({\bf x}-{\bf y}) -\frac{\delta{\cal G}^{(0)}({\bf y})}{\delta\eta}\right)=0 \nonumber \\
\ea
with the solution
\be\label{21}
{\cal G}^{(0)} =\left( -\partial^{i}F_{ij} -{\xi} A_j \right)\partial^{j}\eta +\left(-\partial^{i}\pi_{i} -{\xi} A_{0} \right)\partial^{0}\eta.
\ee

With this result we have a new symplectic potential,
\be\label{22}
\tilde{V}^{(1)}=\tilde{V}^{(0)} + {\cal G}^{(0)} + \sum_{n=1}^{\infty} {\cal G}^{(n)}
\ee
and again, a contraction of this result with the zero-mode give us the next correction to the function $G$:
\ba\label{23}
& &\int dy \left( -{\xi} \partial_{y}^{i}\eta({\bf y})\partial^{x}_{i}\delta({\bf x}-{\bf y}) \right. \nonumber \\
&-& \left. {\xi} \partial_{y}^{0}\eta({\bf y})\partial^{x}_{0}\delta({\bf x}-{\bf y})-\frac{\delta{\cal G}^{(1)}({\bf y})}{\delta\eta}\right)=0,
\ea
namely,
\be\label{24}
{\cal G}^{(1)}=-\frac{\xi}{2}\partial_i \eta \partial^i \eta - \frac{\xi}{2}\partial_0 \eta \partial^0 \eta .
\ee

Note that the second-order correction term has dependence only on the WZ field, thus all the correction terms ${\cal G}^{(n)}$ for $n\geq 2$ are zero. Then, the gauge invariant first-order Lagrangian density, after some algebra, is given by
\be\label{25}
\tilde{\cal L}={\cal L}_{MProca} + J_{\mu}\partial^{\mu}\eta +\frac{\xi}{2}\partial_\mu \eta \partial^\mu \eta
\ee
where ${\cal L}_{MProca}$ is given in (\ref{01}) and 
\be\label{26}
J_{\mu} = \partial^{\nu}F_{\nu\mu} +{\xi} A_{\mu}
\ee
is the Euler current.

For convenience, let us define $\partial_\mu \eta$ as an external field $B_\mu$, and we can rewrite (\ref{25}) as
\be\label{27}
\tilde{\cal L}={\cal L}_{MProca} + J_{\mu}B^{\mu} +\frac{\xi}{2}B_\mu B^\mu \,\,,
\ee
and solving the equation of motion for $B_\mu$ we have,
\be\label{28}
\tilde{\cal L}={\cal L}_{MProca} -\frac{1}{2\xi}J_\mu J^\mu.
\ee

Finally, eliminating the WZ fields we can work algebraically  to obtain the final equivalent  theory as
\be\label{29}
\tilde{\cal L}=-\frac{1}{4}F_{\mu\nu}F^{\mu\nu} -\frac{1}{2\xi}(\partial_\lambda F^{\mu\nu})(\partial^{\lambda}F_{\mu\nu}),
\ee
which is the Maxwell-Podolsky model \cite{Podol}, where the dimension of $\xi$ is $[mass]^2$.

We found that, starting from the Maxwell-Proca model, where the gauge symmetry is not present, and using  the symplectic embedding method we found the Maxwell-Podolsky model that propagates a massive photon and preserves the gauge symmetry.

As well known from the symplectic formalism literature, the zero-mode is the generator of the infinitesimal gauge transformation given by $\delta{\cal O}=\epsilon \tilde{\nu}^{(0)}$. So, using the zero-mode given in (\ref{14}), we have the following gauge symmetries,
\ba\label{30}
\delta A_i &=& -\partial_i \epsilon \nonumber\\
\delta \pi_i &=& 0\nonumber\\
\delta A_0 &=& -\partial_0 \epsilon \nonumber\\
\delta \eta &=& \epsilon,
\ea
where $\epsilon$ is an infinitesimal time-dependent parameter.

The gauge transformations obtained above completes the process of symplectic embedding which is altogether different from the gauge embedding technique, as stressed above.  We believe that the arbitrariness of the zero-mode choice in order to obtain a whole family of dual partners is a positive factor when we compare both procedures.  

It is important to notice the dual aspects of the electrodynamics that appear in the Proca and Podolsky models.  This duality is also important in the extension of the bosonization programme from $D=2$ to higher dimensions \cite{Bazeia}.

\section{The spectrum analysis of the Maxwell-Podolsky theory}\label{s4}

In this section, we derive the propagators of the Maxwell-Podolsky theory in order to carry out a study about its spectrum.

After a straightforward computation, we can  write the Lagrangian (\ref{29}) as
\be\label{31}
{\cal L} = \frac{1}{2}A^{\mu}\left( g_{\mu\nu}{\Box} - \partial_{\mu}\,\partial_{\nu}\,-\,\frac{1}{\xi}g_{\mu\nu}{\Box}^2 + \frac{1}{\xi}{\Box} \partial_\mu \partial_\nu \right)A^{\nu}.
\ee
Let us rewrite (\ref{31}) conveniently as,
\be\label{32}
{\cal L}=\frac{1}{2}A^{\mu} \Theta_{\mu\nu}A^{\nu},
\ee
where
\be\label{33}
\Theta_{\mu\nu}=g_{\mu\nu}{\Box}\,-\,\partial_{\mu}\,\partial_{\nu}\,-\,\frac{1}{\xi}g_{\mu\nu}{\Box}^2 + \frac{1}{\xi}{\Box} \partial_\mu \partial_\nu 
\ee
is a differential operator to the theory with $g_{\mu\nu}=\theta_{\mu\nu}+\omega_{\mu\nu}$. The $\theta_{\mu\nu}$ is the transversal projector and $\omega_{\mu\nu}=\frac{\partial_\mu \partial_\nu }{\Box}$ is the longitudinal projector.

The inverse operator can then be written as
\be\label{34}
\Theta_{\mu\nu}^{-1} =\frac{1}{\Box(1-\xi^{-1} \Box )}\theta_{\mu\nu},
\ee
where the vector propagator is defined as 
\be\label{35}
\left\langle A_\mu A_\nu \right\rangle =i\Theta_{\mu\nu}^{-1}=\frac{1}{\Box(1-\xi^{-1} \Box )}\theta_{\mu\nu}.
\ee
In momentum space the (\ref{35}) is expressed by the relation
\be\label{36}
\left\langle A_\mu A_\nu \right\rangle =\frac{-i}{k^2(1-\xi^{-1}k^2 )}\tilde{\theta}_{\mu\nu},
\ee
or
\be\label{37}
\left\langle A_\mu A_\nu \right\rangle =\frac{-i}{k^2}\tilde{\theta}_{\mu\nu}-\frac{i\xi^{-1}}{1-\xi^{-1}k^2 }\tilde{\theta}_{\mu\nu},
\ee
The relation (\ref{37}) is the vector propagator for the Maxwell theory extended due to the Podolsky term in the (\ref{29}). Analyzing the dispersion relations corresponding to the propagator of the Maxwell-Podolsky, (\ref{29}), we find one particle of spin 1 (one massless pole $k^2 =0$) and other particle of spin 1 with a massive pole $k^2=-\frac{1}{\xi}$. If $\xi < 0$ then $k^2 =\frac{1}{\left|\xi\right|}$ is a massive ``photon'' and it is not a tachyon.

Now, let us to analyze the eigenvalue of the residue matrix for the pole $k^2=-\frac{1}{\xi}$ with $\xi < 0$ , choosing $k^\mu$ purely timelike $k^{\mu} \equiv ( {\left|\xi\right|^{-1/2}} ,\vec{0} )$, we get
\be\label{38}
R_{\mu\nu}=-i\tilde{\theta}_{\mu\nu}=-i\left(g_{\mu\nu} -\frac{k_\mu k_\nu }{k^2} \right)=-i\left(g_{\mu\nu} -\frac{k_\mu k_\nu }{\left|\xi\right|^{-1}} \right)
\ee
or
\be\label{39}
R=
\begin{pmatrix}
0 & 0 & 0 & 0
\cr 0 & i & 0 & 0
\cr 0 & 0 & i & 0
\cr 0 & 0 & 0 & i
\end{pmatrix}
\ee 
We calculate its eigenvalues and find three nonvanishing eigenvalues
\be\label{40}
\lambda =
\begin{pmatrix}
i & i & i
\end{pmatrix}\,\,.
\ee 
Hence, for timelike $k^\mu$, the poles of $\left\langle A_\mu A_\nu \right\rangle$ respect causality (they are not tachyonic) and correspond to physically acceptable one-particle states with three degree of freedom, since the residue matrix exhibits a single positive eigenvalue. Consequently, the model may be adopted as a consistent theory.

\section{Final discussions}\label{s5}

In this work we used the symplectic embedding procedure to promote the dualization of the gauge invariant Maxwell theory in $D=4$, modified by the introduction of an explicit massive (Proca) term.  This technique was introduced originally to deal with constrained models and involves the construction of a symplectic tensor and a zero-mode term.

The new theory obtained (the dual partner) involves a massive term for the vector field, which is the Podolsky term that involves higher-order derivatives and propagates a massive ``photon'' without a spontaneous breaking of the gauge symmetry. 

In comparison with the gauge embedding formalism, although being a swift iterative technique in order to obtain a dual partner of the original theory, a question about the uniqueness lingers.  To use the symplectic embedding technique means to obtain a whole family of dual partners thanks to the arbitrariness of the zero-mode term.  Besides, the zero-mode term is the generator of the infinitesimal gauge transformations of the dual partner theory.

The duality between the Proca and the Podolsky theories is also important in the extension of the bosonization programme from $D=2$ to higher dimensions.  Although out of the scope of this work, we can say that bosonization is very important for the interpretation of the new parameters in the Podolsky extension.

To infer about the physical nature of the simple poles, we have to calculate the eigenvalues of the residue matrix for each of these poles.  To carry out this investigation we analyzed the pole structure of the vector propagator by the residue matrix, where we found three positive eigenvalues that correspond to the physically acceptable one-particle states with three degree of freedom.  When $\xi \rightarrow 0$, we easily obtain the original QED (massless photon).

\section{ Acknowledgments}
The authors would like to thank CNPq, FAPEMIG and FAPERJ (Brazilian financial agencies) for financial support and to thank I. C. S. Barroca for the critical reading.

\end{document}